\documentclass[a4paper]{jpconf}
\usepackage{graphicx}
\usepackage{amsmath,amssymb,amsfonts,epsf}

\def\nn{\nonumber}

\newcommand{\be}{\begin{equation}}
\newcommand{\ee}{\end{equation}}
\def\ba{\begin{array}}
\def\ea{\end{array}}
\def\ft#1#2{{\textstyle{\frac{\scriptstyle #1}{\scriptstyle #2}}}}
\def\fft#1#2{\frac{#1}{#2}}

\newcommand{\bea}{\begin{eqnarray}}
\newcommand{\eea}{\end{eqnarray}}

\def\R{\rlap{\rm I}\mkern3mu{\rm R}}
\begin{document}

\title{The Behavior of Strings on AdS wormholes}

\author{Mir Ali$^1$, Frenny Ruiz$^1$, Carlos Saint-Victor$^1$\\ and Justin F V\'azquez-Poritz$^{1,2}$}

\address{$^1$ Physics Department, New York City College of Technology\\ The City University of New York, 300 Jay Street, Brooklyn NY 11201, USA}
\address{$^2$ The Graduate School and University Center, The City University of New York\\ 365 Fifth Avenue, New York NY 10016, USA}

\ead{jvazquez-poritz@citytech.cuny.edu}

\begin{abstract}
We consider the behavior of open strings on AdS wormholes in Gauss-Bonnet theory, which are the Gauss-Bonnet gravity duals of a pair of field theories. A string with both endpoints on the same side of the wormhole describes two charges within the same field theory, which exhibit Coulomb interaction for small separation. On the other hand, a string extending through the wormhole describes two charges which live in different field theories, and they exhibit a spring-like confining potential. A transition occurs when there is a pair of charges present within each field theory: for small separation each pair of charges exhibits Coulomb interaction, while for large separation the charges in the different field theories pair up and exhibit confinement. If two charges move faster than a critical speed, then they exhibit a separation gap and energy is transferred from the leading charge to the lagging one. 

To appear in the proceedings of Quantum Theory and Symmetries 6, held at the University of Kentucky on July 20-25 2009.
\end{abstract}

\section{Introduction and summary}

The AdS/CFT correspondence \cite{ads4} can be used to study certain strongly-coupled gauge theories. In particular, the behavior of open strings can be related to that of particles in the field theory. In the large $N$ and large 't Hooft coupling $\lambda$ limit, the string theory can be approximated by classical supergravity. A quantum theory of gravity such as string theory generally contains higher derivative corrections from stringy or quantum effects, which correspond to $1/\lambda$ or $1/N$ corrections in the field theory. However, not much is known about the precise forms of the higher derivative corrections, other than for a few maximally supersymmetric cases. The most general theory of gravity in higher dimensions that leads to second-order field equations for the metric is described by the Lovelock action \cite{lovelock}. The simplest such higher derivative theory of gravity is known as Einstein-Gauss-Bonnet theory, which only contains terms up to quadratic order in the curvature. Our primary motivation for considering Einstein-Gauss-Bonnet theory is that it contains Lorentzian-signature five-dimensional asymptotically locally AdS wormhole solutions which do not violate the weak energy condition, so long as the Gauss-Bonnet coupling constant is negative and bounded according to the shape of the solution \cite{gbwormholes,maeda}. 

We consider the static wormhole solutions that were found in \cite{troncoso1,troncoso2} which connect two asymptotically AdS spacetimes and are described by the metric
%%%%
\be\label{metric1}
ds^2=\ell^2 \Big(-\cosh^2(\rho-\rho_0) dt^2+d\rho^2+\cosh^2\rho\ d\Sigma_3^2\Big)\,.
\ee
%%%%
The base manifold $\Sigma_3$ can be either $H_3/\Gamma$ or $S^1\times H_2/\Gamma$, where $H_3$ and $H_2$ are hyperbolic spaces and $\Gamma$ is a freely acting discrete subgroup of $O(3,1)$ and $O(2,1)$, respectively. The hyperbolic part of the base manifold must be quotiented so that this geometry describes a wormhole rather than a gravitational soliton with a single conformal boundary.
The neck of the wormhole is located at $\rho=0$, which connects two asymptotically locally AdS regions at $\rho\rightarrow\pm\infty$. The geometry is devoid of horizons and radial null geodesics connect the two asymptotic regions in a finite time $\Delta t=\pi$. These wormholes can evade the proof that the disconnected boundaries must be separated by black hole horizons \cite{theorem}, since we are dealing with a higher-derivative theory of gravity. No energy conditions are violated by these wormholes, since the stress-energy tensor vanishes everywhere \cite{troncoso1,troncoso2}. The stability of these wormholes against scalar field perturbations has been discussed in \cite{stability}.

It has been shown that gravity pulls towards a fixed hypersurface at $\rho=\rho_0$, which lies in parallel with the neck of the wormhole \cite{troncoso1,troncoso2}. In particular, timelike geodesics are confined, and oscillate about this hypersurface. Although the wormhole is massless, for nonzero $\rho_0$, the mass of the wormhole appears to be positive for observers located at on one side and negative for the other. The parameter $\rho_0$ is related to the apparent mass on each side of the wormhole; for $\rho_0=0$, the wormhole exhibits reflection symmetry. 

If we can apply the AdS/CFT correspondence to these AdS wormholes, then this is the gravity dual of two interacting field theories on $\R\times \Sigma_3$. Except in the UV limit, the conformal symmetry of both theories is broken by the length scale associated with $\Sigma_3$ ($H_3$ and $H_3$ must have the radius $1$ and $1/\sqrt{3}$, respectively) as well as by the parameter $\rho_0$. However, we would like to emphasize that the AdS/CFT correspondence has not actually been tested for gravitational backgrounds of Einstein-Gauss-Bonnet theory and may not be valid. We are making the working assumption that the low-energy effective five-dimensional description of gauge theory/string theory duality has a sensible derivative expansion in which the higher curvature terms are systematically suppressed by powers of the Planck length. After making the appropriate field redefinitions, the curvature-squared terms in the action appear in the Gauss-Bonnet combination. Thus, within the limitations of the derivative expansion, one can use the AdS/CFT correspondence to determine the properties of gauge theories dual to Einstein-Gauss-Bonnet gravitational backgrounds \cite{buchel}. In fact, one might expect that within the vast string landscape there are higher derivative corrections which lead to similar backgrounds with asymptotically AdS regions with multiple disconnected boundaries. 

We consider the behavior of open strings on these AdS wormholes. The string endpoints lie on probe D-branes and correspond to charges in one or the other of the field theories. Technically, we should also consider the quantum fluctuations of the string worldsheet \cite{drukker}, which formally dominate over the higher derivative corrections of the background. However, we have checked that the corrections to the string worldsheet do not change the qualitative behavior of strings, which is the focus of our interest. 

The expectation value of a rectangular Wilson loop can be computed from the proper area of the string worldsheet \cite{wilson1,wilson2}, from which the energy of a pair of charges can be read off. We generalize this prescription for the case of two charges which do not live within the same field theory. Even though the string endpoints lie on opposite sides of the wormhole, there is a rectangular contour within the physical spacetime whose horizontal sides run along the distance between the two charges and whose vertical sides run along time. We find that two charges in different field theories exhibit confinement with a spring-like potential. A pair of heavy charges exhibits a weaker interaction than do lighter charges. We find that there is a rather curious transition which involves a quadruplet of charges that consists of a charge-anticharge pair within each field theory. For small charge-anticharge separations, each pair exhibits Coulomb interaction and does not interact with the pair of charges in the other field theory. However, for large separation, it is the charges of different types that interact and exhibit confinement whereas the pairs of charges of the same type no longer interact with each other. In this sense, each pair of charges of the same type exhibits the feature of effectively having a screening length, even though both field theories are at zero temperature. 

We also consider the behavior of steadily-moving strings. In particular, demanding that the string configuration is timelike imposes constraints on the mass parameters of the corresponding charges, as well as on their separation. We find that the charges have an upper bound on their speed which depends on the mass parameter of one of the charges. This speed limit is generally less than the speed of light, which is a result of the fact that the proper velocity of the string endpoints is greater than the physical velocity in the field theory \cite{argyres}. We find that a pair of steadily-moving charges of different types can occupy the same location with no transfer of energy or momentum between them, provided that their speed is less than a certain critical speed. This critical speed coincides with the speed limit of the charges and monotonically decreases with the mass parameters. If the charges move at this critical speed, then they exhibit a separation gap which increases with the mass parameter of the lagging charge but is not so sensitive to the mass parameter of the leading charge. Also at this critical speed, energy and momentum are transferred from the leading charge to the lagging one.

\section{Static strings}

We will consider a string which is localized at a point in $H_2$ and moves along the $S^1$ direction, which we label $x$. Choosing a static gauge, we find that the equation of motion for a string on the background metric (\ref{metric1}) is given by
%%%%
\be\label{eom}
\fft{\partial}{\partial\rho} \Big( \fft{\cosh^2(\rho-\rho_0) \cosh^2\rho\ x^{\prime}}{\sqrt{-g}}\Big)-\cosh^2\rho\ \fft{\partial}{\partial t} \Big( \fft{\dot x}{\sqrt{-g}}\Big)=0\,,
\ee
%%%%
where
%%%%
\be\label{g}
-\fft{g}{\ell^4}=\cosh^2(\rho-\rho_0)+\cosh^2(\rho-\rho_0) \cosh^2\rho\ x^{\prime 2}-\cosh^2\rho\ \dot x^2\,.
\ee
%%%%
Note that $g=\mbox{det} (g_{ab})$, where $g_{ab}$ is the induced metric. We will also make use of the fact that the string tension is given by $T_0=\sqrt{\lambda}/(2\pi \ell^2)$, where $\lambda$ is the 't Hooft coupling of the field theory.

The simplest solution is a constant $x=x_0$, for which the string is straight, as shown in Figure 1. Suppose that the endpoints are located at $\rho=-\rho_1$ and $\rho=\rho_2$, where we will always take $\rho_1, \rho_2>0$. We will refer to the charges associated with the endpoints at $\rho=-\rho_1$ and $\rho=\rho_2$ as a type 1 charge (blue) and a type 2 charge (red), respectively. Also, we will refer to the  radial locations of the string endpoints $\rho_1$ and $\rho_2$ as the mass parameters of the charges.
The D-branes (or at least the part that lies closest to the neck of the wormhole) are denoted by the aqua-colored surfaces and the grey surface represents the fixed hypersurface at $\rho=\rho_0$.
%%%%
\begin{figure}[ht]
   \epsfxsize=2.1in \centerline{\epsffile{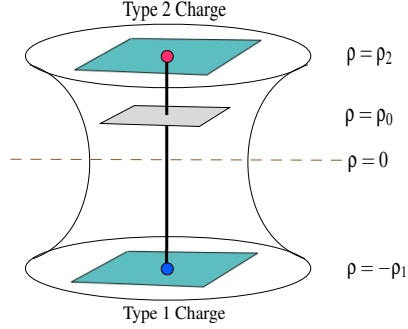}}
   \caption[FIG. \arabic{figure}.]{A static string stretching straight through a wormhole.}
\label{fig1}
\end{figure}
%%%%
The static string has vanishing momentum, and its energy is given by
%%%%
\be\label{restE}
E=T_0 \ell^2\big( \sinh(\rho_1+\rho_0)+\sinh (\rho_2-\rho_0)\big)\,.
\ee
%%%%

\begin{figure}[ht]
\begin{center}
$\begin{array}{c@{\hspace{1.25in}}c}
\epsfxsize=1.4in \epsffile{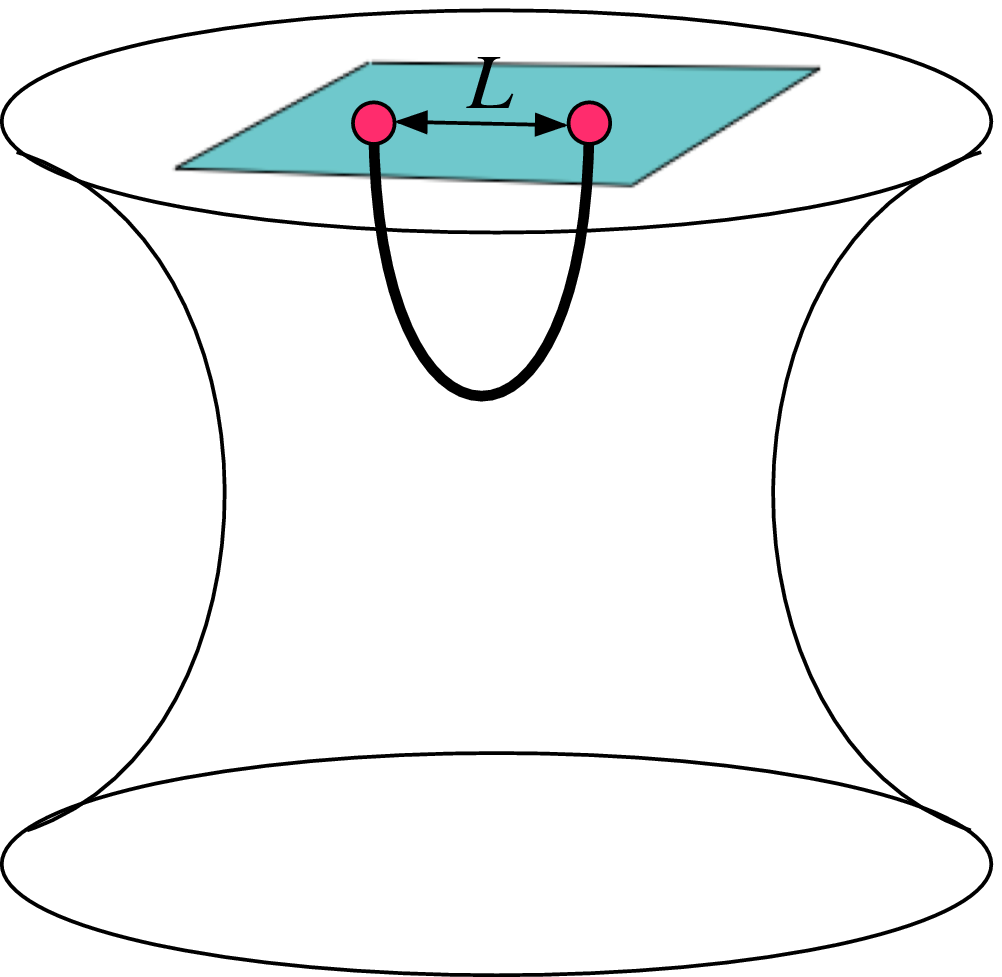} &
\epsfxsize=1.4in \epsffile{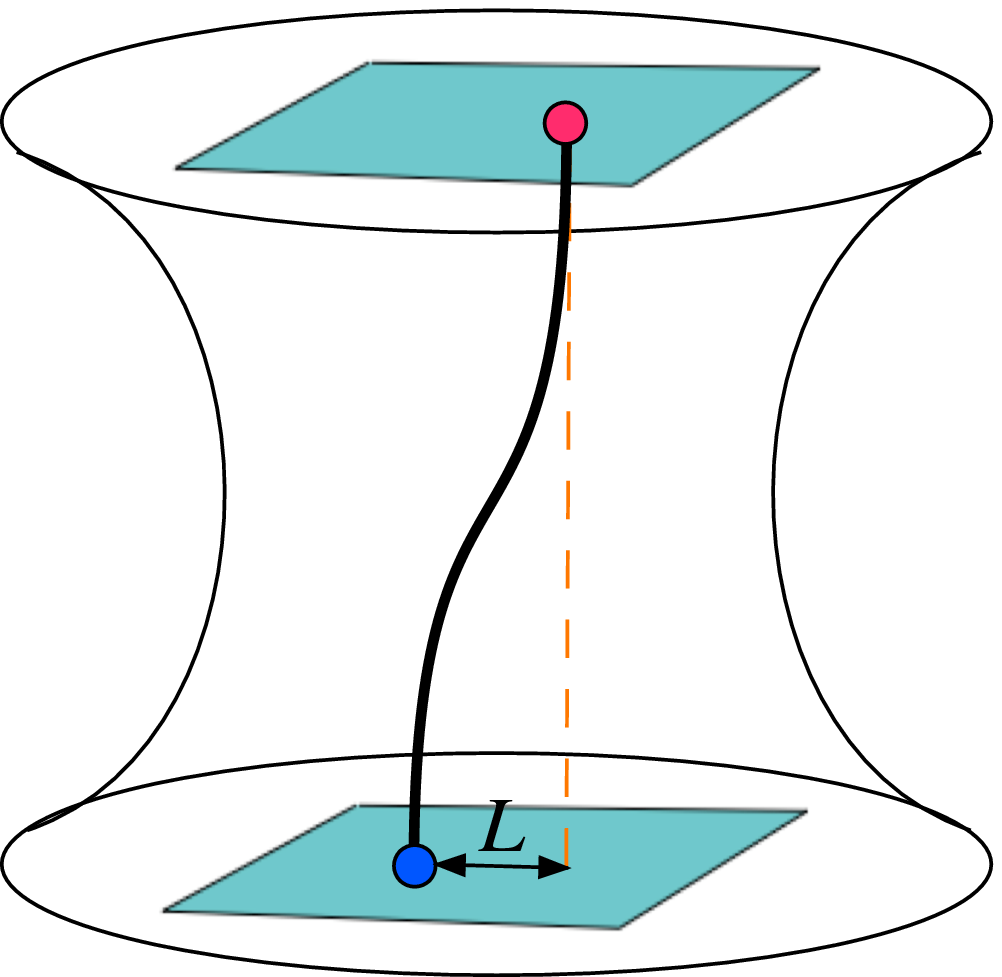}
\end{array}$
\end{center}
\caption[FIG. \arabic{figure}.]{A string with both endpoints on the same side of the wormhole (left) and on opposite sides (right).}
   \label{fig4}
\end{figure}

Curved strings can either have both endpoints on the same side of the wormhole or else on opposite sides, as shown in Figure \ref{fig4}. 
%%%%
%%%%
We will first consider the first case, which corresponds to a pair of charges of the same type. This means that the string dips down towards the wormhole and then comes back up. We will denote the turning point of the string by $\rho_t$, and $L$ is the distance between the endpoints of the string. For simplicity, we will take the endpoints to be at $\rho=\rho_2$, so that we have a pair of type 2 charges. For $\rho_2$, $\rho_t\gg 1$, the string is not sensitive to the presence of the wormhole and so it can be found that its energy goes as $1/L$. A Coulomb potential energy is consistent with the fact that the theory is conformal in the extreme UV regime. If the masses associated with the charges are decreased, or if the distance between the charges is increased, then the string dips closer to the wormhole. Then terms in the energy that are of higher order in $L$ will become important.

From (\ref{eom}) and (\ref{g}), we find the distance $L$ between the string endpoints is given by
%%%%
\be
L=2\cosh (\rho_t-\rho_0) \cosh\rho_t \int_{\rho_t}^{\rho_2} \fft{d\rho}{\cosh\rho \sqrt{\cosh^2 (\rho-\rho_0) \cosh^2\rho-\cosh^2 (\rho_t-\rho_0)\cosh^2\rho_t}}\,.
\ee
%%%%
The above integration region assumes that $\rho_2>\rho_t$. As we will see shortly, there are string solutions for which $\rho_2<\rho_t$, in which case the upper and lower integration bounds must be interchanged.

\begin{figure}[ht]
\begin{center}
$\begin{array}{c@{\hspace{1.0in}}c}
\epsfxsize=2.0in \epsffile{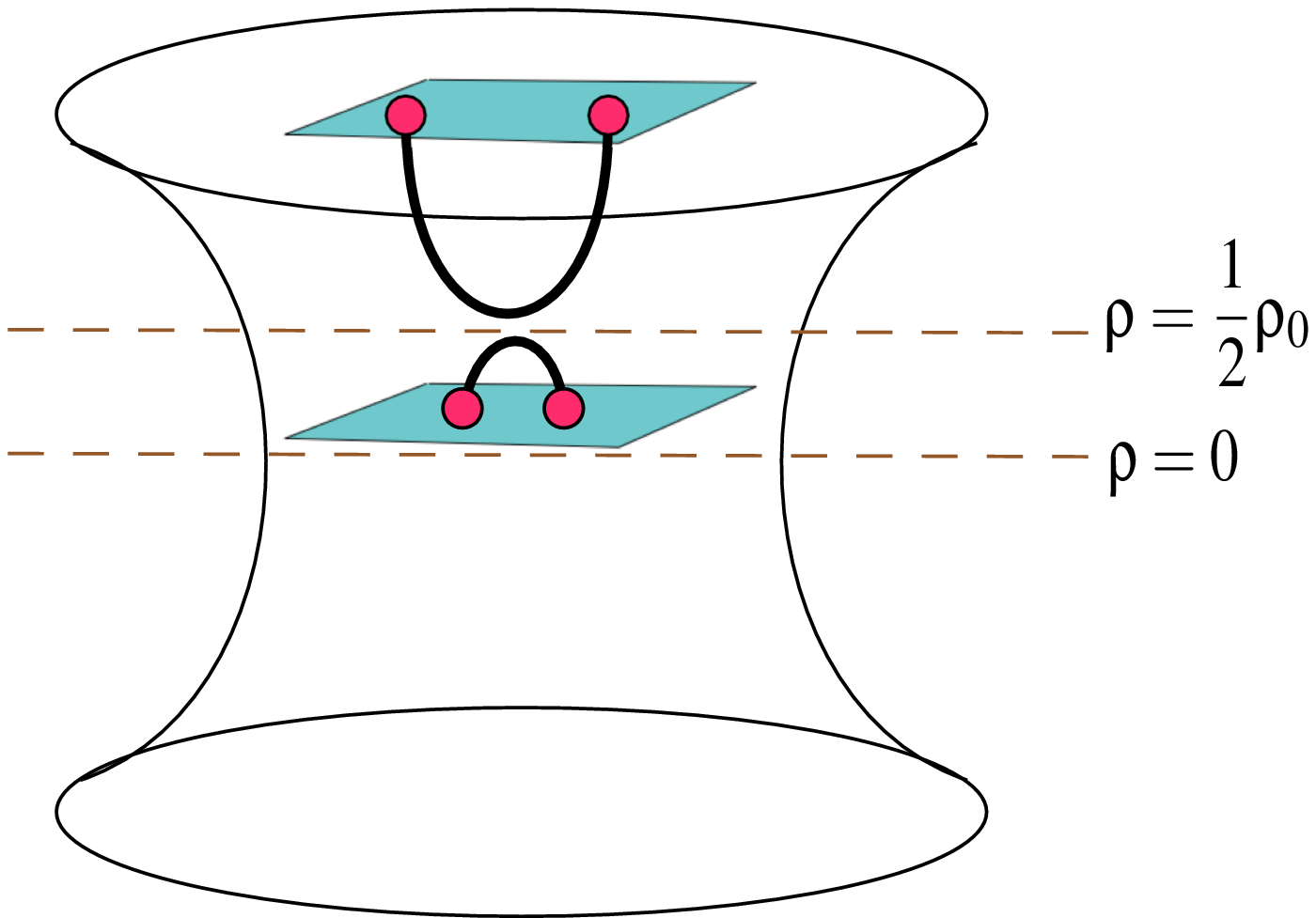} &
\epsfxsize=2.0in \epsffile{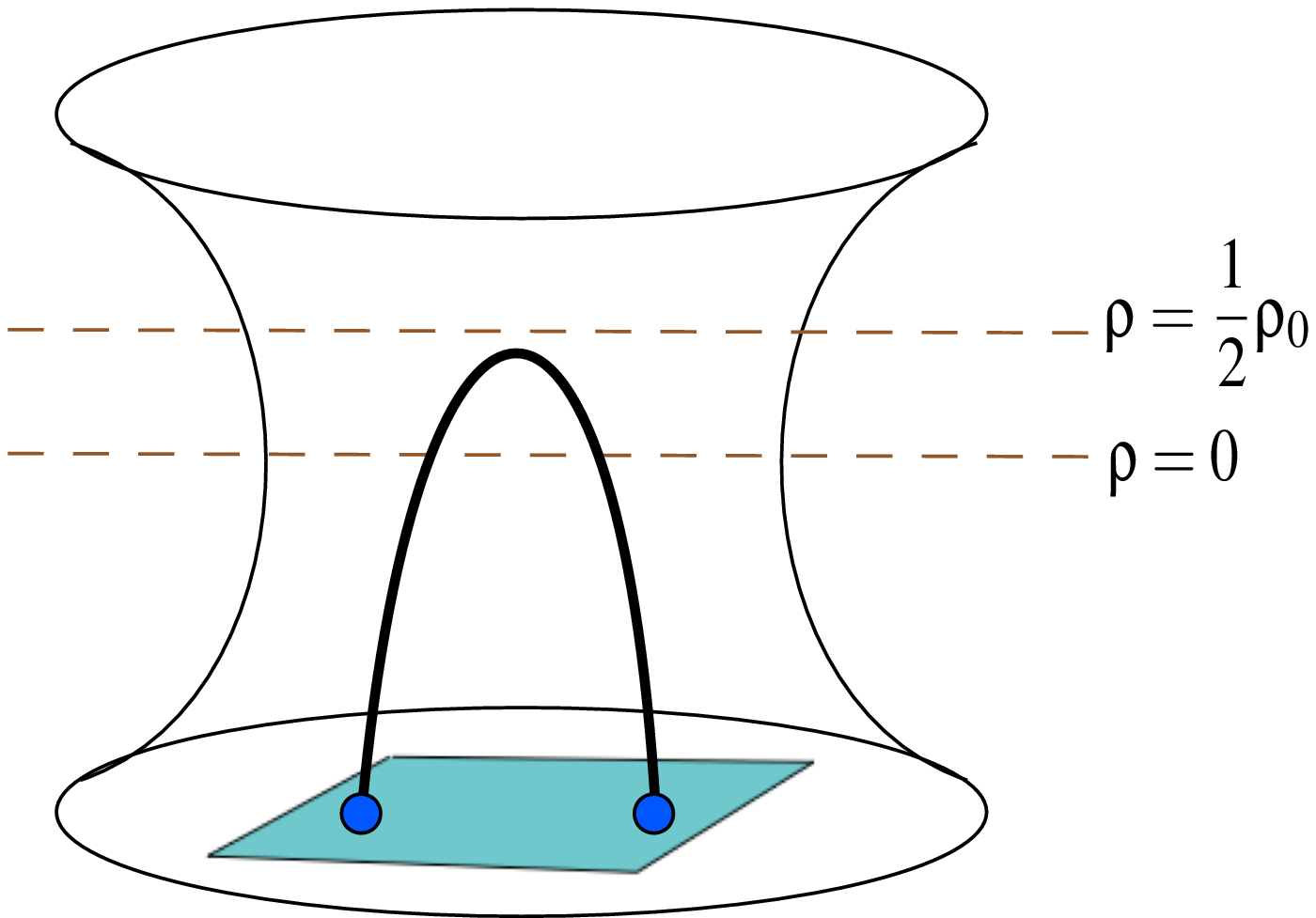}
\end{array}$
\end{center}
\caption[FIG. \arabic{figure}.]{Strings bend towards $\rho=\ft12\rho_0$ regardless of where the string endpoints are located, even if this means that the string bends away from the neck of the wormhole (left) or goes through the neck of the wormhole (right).}
   \label{otherside}
\end{figure}

For $\rho_2>\ft12 \rho_0$, the reality of $L$ implies that the turning point is located in the interval $\rho_2>\rho_t\ge\ft12 \rho_0$. Similarly, a string with endpoints at $\rho_2<\ft12\rho_0$ has a turning point in the interval $\rho_2<\rho_t\le\ft12 \rho_0$. This means that if the turning points are within the interval $0<\rho_2<\rho_0/2$, then the string will bend {\it away} from the center of the wormhole. Moreover, if the endpoints are located on the opposite side of the wormhole as the hypersurface $\rho=\rho_0$, then there are strings which go through the neck of the wormhole and have a midsection on the opposite side of the wormhole as its endpoints. All of these possibilities are shown in Figure \ref{otherside}. 
%%%%
%%%%

The turning point $\rho_t$ monotonically decreases with $L$. In particular, as $L$ gets large the turning point asymptotically approaches $\ft12 \rho_0$ from above. Note that, unlike the case of strings on an AdS black hole background, there is a single string configuration for each value of $L$. From the field theory perspective, this means that the charges do not exhibit a screening length but rather remain interacting no matter how far they are from each other. As the distance between the charges increases, the inter-charge potential becomes ever more sensitive to the IR characteristics of that sector of the theory. 

We will now consider a curved string with endpoints on opposite sides of the wormhole, which corresponds to a type 1 charge and a type 2 charge. From (\ref{eom}) and (\ref{g}), the distance between the string endpoints in the $x$ direction is
%%%%
\be
L=C \int_{-\rho_1}^{\rho_2} \fft{d\rho}{\cosh\rho\sqrt{\cosh^2(\rho-\rho_0)\cosh^2\rho-C^2}}\,.
\ee
%%%%
For small separation $L$, we can expand in $C\ll 1$ so that
%%%%
\be\label{LC}
L\approx C \int_{-\rho_1}^{\rho_2} \fft{d\rho}{\cosh (\rho-\rho_0) \cosh^2\rho}\,.
\ee
%%%%
We find the energy of the string to be
%%%%
\be
E=T_0\ell^2 \int_{-\rho_1}^{\rho_2} d\rho\ \fft{\cosh^2(\rho-\rho_0)\cosh\rho}{\sqrt{\cosh^2(\rho-\rho_0)\cosh^2\rho-C^2}}\,.
\ee
%%%%
In the small separation limit $C\ll 1$,
%%%%
\be
E=E_{straight}+\fft12 T_0\ell^2 C^2 \int_{-\rho_1}^{\rho_2} \fft{d\rho}{\cosh (\rho-\rho_0)\cosh^2\rho}\,,
\ee
%%%%
where $E_{straight}$ is the energy of a straight string given by (\ref{restE}). Using (\ref{LC}) to express $C$ in terms of $L$, we find
%%%%
\be
E=E_{straight}+\fft12 kL^2\,,
\ee
%%%%
where
%%%%
\be
k=\fft{\sqrt{\lambda}}{2\pi} \Big( \int_{-\rho_1}^{\rho_2} \fft{d\rho}{\cosh (\rho-\rho_0)\cosh^2\rho}\Big)^{-1}\,,
\ee
%%%%
and we have used $T_0=\sqrt{\lambda}/(2\pi\ell^2)$. Thus, for small $L$, a pair of charges of opposite types are confined and exhibit a spring-like potential, as opposed to a pair of charges of the same type which exhibit a Coulomb potential. Since we are at zero temperature, there is no screening length. The effective force constant $k$ monotonically decreases with the mass parameters $\rho_1$ and $\rho_2$, which means that heavy charges are not as sensitive to the confining potential as lighter charges. 

\begin{figure}[ht]
\begin{center}
$\begin{array}{c@{\hspace{1.25in}}c}
\epsfxsize=1.4in \epsffile{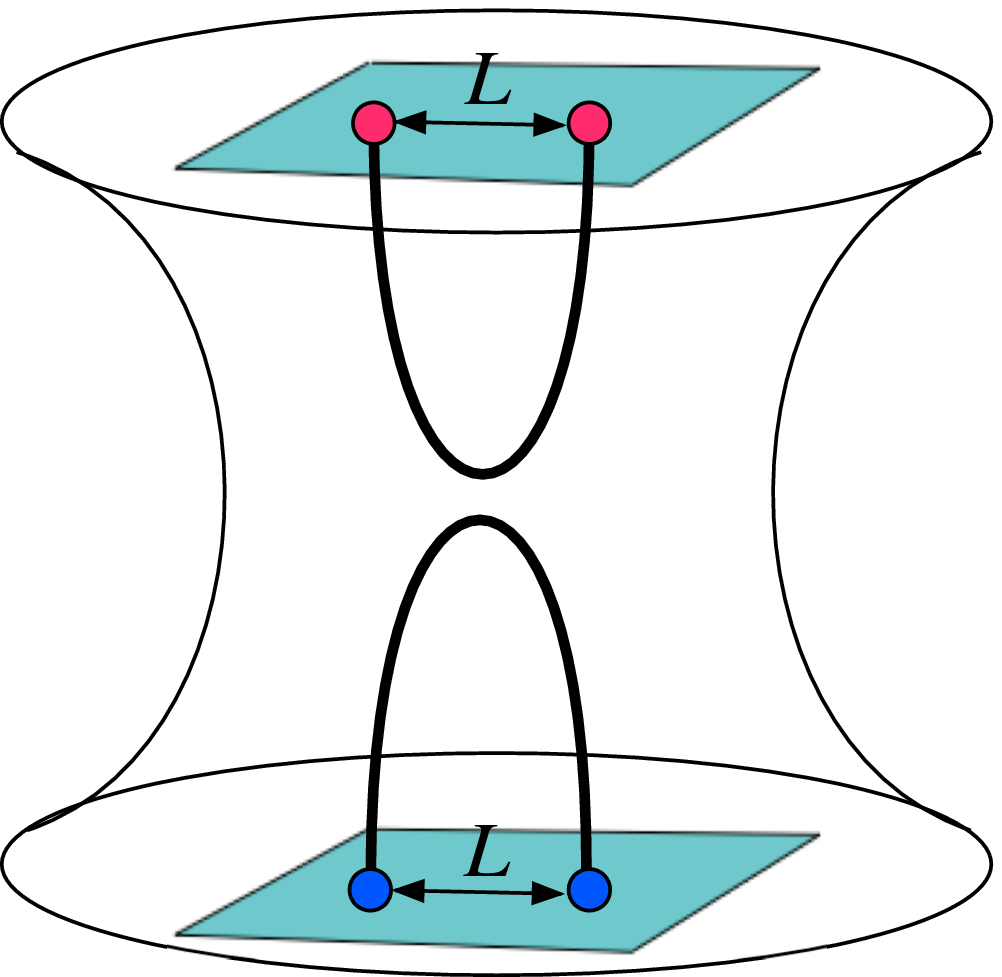} &
\epsfxsize=1.4in \epsffile{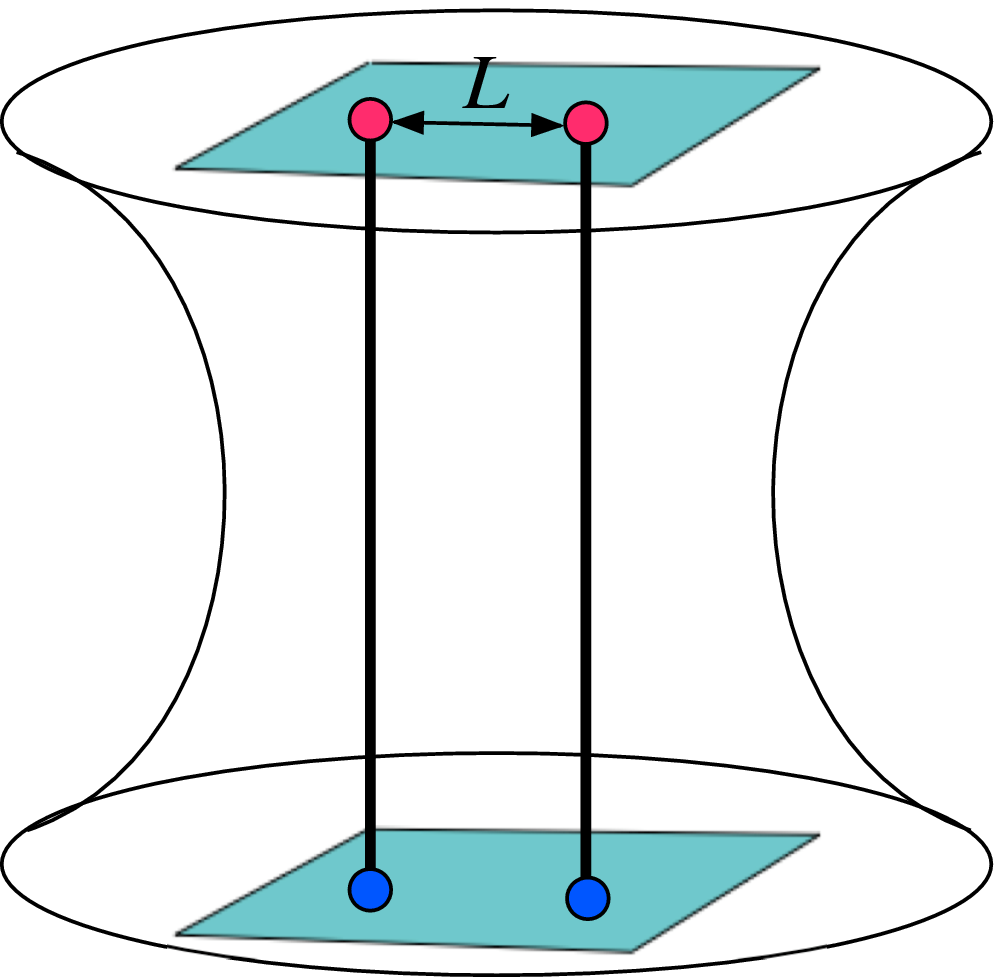}
\end{array}$
\end{center}
\caption[FIG. \arabic{figure}.]{There is a critical distance $L_{crit}$ associated with two sets of charge-anticharge pairs. For $L<L_{crit}$, the left configuration is energetically favorable while, for $L>L_{crit}$, the right configuration is the favored one.}
   \label{figtrans}
\end{figure}

For two sets of charge-anticharge pairs, an interesting transition can take place. For instance, suppose we have a pair of interacting type 1 charges and a pair of interacting type 2 charges, described by the two curved strings shown in the left plot of Figure \ref{figtrans}. For simplicity, we will assume that both pairs of charges are a distance $L$ apart. For small $L$, there is no interaction between the type 1 pair and the type 2 pair. However, as $L$ increases, the strings bend closer to each other. Note that these strings cannot pass each other, since neither one can pass the radius $\rho=\rho_0/2$. However, even before either string hits this bound, the string configuration in the right plot of Figure \ref{figtrans} becomes the energetically favorable one with the same set of endpoints. Thus, for large $L$, each type 1 charge becomes coupled to a type 2 charge. This is similar to a screening length, in that the charges of the same type no longer interact with each other. However, since we are at zero temperature, this is really a feature of having multiple pairs of charges close to each other.
%%%%
%%%%

\section{Steadily-moving strings}

\subsection{Straight strings}

A steadily-moving straight string is given by $x(t,\rho)=x_0+vt$. For this solution, we find that
%%%%
\be
-\fft{g}{\ell^4}=\cosh^2(\rho-\rho_0)-v^2 \cosh^2\rho\,.
\ee
%%%%
For simplicity, we will take $\rho_0\ge 0$. Then for $v< v_{crit}\equiv e^{-\rho_0}$, $-g>0$ everywhere for $v<1$, which implies that no parts of the string move faster than the local speed of light. On the other hand, for $v=v_{crit}$, we require that $\rho_2<\rho_0/2$. For $v> v_{crit}$, $g=0$ at $\rho=\rho_{crit}$ where
%%%%
\be\label{pcrit}
\rho_{crit}=\ln \sqrt{\fft{e^{\rho_0}-v}{v-e^{-\rho_0}}}\,,
\ee
%%%%
at which point the induced metric on the string worldsheet is degenerate. We can still have steadily-moving straight strings which lie entirely within the region $\rho_{crit}<\rho$. However, for strings which have a region that lies within $\rho<\rho_{crit}$, $-g<0$ and the action, energy and momentum are complex. This is a signal that this part of the string travels faster than the local speed of light and must be discarded. 

The energy and momentum of the steadily moving string are given by the integrals
%%%%
\bea\label{Ep}
E &=& T_0\ell^2 \int_{-\rho_1}^{\rho_2} d\rho\ \fft{\cosh^2 (\rho-\rho_0)}{\sqrt{\cosh^2(\rho-\rho_0)-v^2 \cosh^2\rho}}\,,\nn\\ 
p &=& T_0\ell^2v \int_{-\rho_1}^{\rho_2} d\rho\ \fft{\cosh^2 \rho}{\sqrt{\cosh^2(\rho-\rho_0)-v^2 \cosh^2\rho}}\,.
\eea
%%%% 
For the special case of $\rho_0=0$, we find that 
%%%%
\be
E=\fft{E_{rest}}{\sqrt{1-v^2}}\,,\qquad E_{rest}=T_0\ell^2 (\sinh\rho_1+\sinh\rho_2)\,.
\ee
%%%%
For non-vanishing $\rho_0$, the expressions for $E$ and $p$ cannot be integrated in closed form. Note that, since $\pi_t^1$ and $\pi_x^1$ vanish, there is no energy or momentum current flowing along the string.

We will now comment on the speed limit of these co-moving charges, which stems from the fact that the proper velocity $V$ of the string endpoints differs from the physical velocity $v$ in the four-dimensional field theory \cite{argyres}. From the metric (\ref{metric1}), we see that
%%%%
\be
V=\fft{\cosh\rho_i}{\cosh (\rho_i-\rho_0)} v\,,
\ee
%%%%
where $\rho_i=-\rho_1$ or $\rho_2$. In order to avoid a spacelike string worldsheet, we must have $V\le 1$. This corresponds to
%%%%
\be
v\le v_{max}=\fft{\cosh (\rho_2-\rho_0)}{\cosh\rho_2}\,,
\ee
%%%%
where we have taken $\rho_i=\rho_2$, since $\rho_2$ is closer to $\rho_0\ge 0$ than is $-\rho_1$. As is the case for steadily-moving strings in AdS black hole backgrounds, the speed limit depends on the mass parameter associated to the charge. This same speed limit applies to the curved strings which we now discuss. We see that $v_{max}=v_{crit}=1$ for $\rho_0=0$. On the other hand, for $\rho_0>0$, we find that $v_{max}<v_{crit}<1$.

\subsection{Curved strings}

We will now consider steadily-moving curved strings, which are described by solutions of the form $x(t,\rho)=x(\rho)+vt$. The term with the time derivative in (\ref{eom}) vanishes and we are left with
%%%%
\be\label{eom2}
\fft{\partial}{\partial\rho} \Big( \fft{\cosh^2(\rho-\rho_0) \cosh^2\rho\ x^{\prime}}{\sqrt{-g}}\Big)=0\,.
\ee
%%%%
where
%%%%
\be
-\fft{g}{\ell^4}=\cosh^2(\rho-\rho_0)+\cosh^2(\rho-\rho_0) \cosh^2\rho\ x^{\prime 2}-v^2 \cosh^2\rho\,.
\ee
%%%%
The first integral of (\ref{eom2}) is
%%%%
\be
\fft{\cosh^2(\rho-\rho_0) \cosh^2\rho\ x^{\prime}}{\sqrt{-g/\ell^4}}=C\,.
\ee
%%%%
where $C$ is an integration constant. Thus,
%%%%
\be\label{xeqn}
x^{\prime 2}=\fft{C^2}{\cosh^2(\rho-\rho_0)\cosh^2\rho}\ \Big( \fft{\cosh^2(\rho-\rho_0)-v^2 \cosh^2\rho}{\cosh^2(\rho-\rho_0)\cosh^2\rho-C^2}\Big)\,.
\ee
%%%%
Solving for $-g$ gives
%%%%
\be
-\fft{g}{\ell^4}=\cosh^2(\rho-\rho_0)\cosh^2\rho\ \Big( \fft{\cosh^2(\rho-\rho_0)-v^2\cosh^2\rho}{\cosh^2(\rho-\rho_0)\cosh^2\rho-C^2}\Big)\,.
\ee
%%%%
For $v\le v_{crit}$ (with the exception of $\rho_0=0$ and $v=1$) then $-g>0$ everywhere along the string provided that $|C|<C_{crit}$ where
%%%%
\be\label{Ccrit}
C_{crit}=\fft12\big( \cosh (2\mbox{min}[\rho_2,\rho_0/2]-\rho_0)+\cosh\rho_0\big)\,.
\ee
%%%%
In the limit that $\rho_0=0$, solutions exist with $|C|<1$ for all $v<1$. However, for $\rho_0=0$ and $v=1$, the induced metric is degenerate everywhere, regardless of the value of $C$. Strings with $v>v_{crit}$ can still have $-g>0$ everywhere provided that $\rho_2<\rho_{crit}$ and $|C|<C_{crit}$, where $\rho_{crit}$ and $C_{crit}$ are given by (\ref{pcrit}) and (\ref{Ccrit}), respectively. From the field theory perspective, this corresponds to an upper bound on the mass of the type 2 charge. For $v>v_{crit}$, one can also have a string that lives entirely in $\rho>\rho_{crit}$, provided that $|C|>C_{crit}$. This corresponds to a light-heavy pair of type 2 charges with a lower bound on their masses.

The distance between the charges is given by
%%%%
\be
L=C\int_{-\rho_1}^{\rho_2} \fft{d\rho}{\cosh (\rho-\rho_0)\cosh\rho} \sqrt{\fft{\cosh^2 (\rho-\rho_0)-v^2\cosh^2\rho}{\cosh^2 (\rho-\rho_0)\cosh^2\rho-C^2}}\,,
\ee
%%%%
For cases in which we require that $|C|<C_{crit}$, there is an upper bound on $L$. However, if the endpoints of the string are allowed to move freely, then the string will tend to straighten out to the $C=0$ configuration so that it minimizes its energy. On the other hand, for the string corresponding to a light-heavy pair of type 2 charges moving with $v>v_{crit}$, we must have $|C|>C_{crit}$ and there is a corresponding lower bound on $L$. This string cannot straighten out without decreasing its speed.

The rate at which energy and momentum flow through the string can be written in terms of the constant $C$ as
%%%%
\be
\pi_t^1=\fft{\sqrt{\lambda}}{2\pi} Cv\,,\qquad -\pi_x^1=\fft{\sqrt{\lambda}}{2\pi}C\,,
\ee
%%%%
Thus, for nonzero C, energy and momentum are transferred from the leading charge to the lagging one. If there is no external force present, then the leading charge presumably slows down and the lagging charge speeds up until an equilibrium state is reached corresponding to a steadily-moving straight string.

\ack

We would like to thank Philip Argyres, Mohammad Edalati, Hong L\"u and Juan Maldacena for useful correspondence and conversations. M.A., F.R. and C.S.-V. are supported in part by the Emerging Scholars Program. F.R. and C.S.-V. are supported in part by New York City Louis Stokes Alliance for Minority Participation in Science, Mathematics, Engineering and Technology.

\end{document}